# Valley Manipulation by Optically Tuning the Magnetic Proximity Effect in WSe$_2$/CrI$_3$ Heterostructures

Kyle L. Seyler[1], Ding Zhong[1], Bevin Huang[1], Xiayu Linpeng[1], Nathan P. Wilson[1], Takashi Taniguchi[2], Kenji Watanabe[2], Wang Yao[3], Di Xiao[4], Michael A. McGuire[5], Kai-Mei C. Fu[1,6], Xiaodong Xu[1,7*]

[1]Department of Physics, University of Washington, Seattle, Washington 98195, USA
[2]National Institute for Materials Science, 1-1 Namiki, Tsukuba 305-0044, Japan
[3]Department of Physics and Center of Theoretical and Computational Physics, University of Hong Kong, Hong Kong, China
[4]Department of Physics, Carnegie Mellon University, Pittsburg, Pennsylvania 15213, USA
[5]Materials Science and Technology Division, Oak Ridge National Laboratory, Oak Ridge, Tennessee, 37831, USA
[6]Department of Electrical Engineering, University of Washington, Seattle, Washington 98195, USA
[7]Department of Materials Science and Engineering, University of Washington, Seattle, Washington 98195, USA

[*]Corresponding author. Email: xuxd@uw.edu, phone: (206) 543-8444

**Abstract:** Monolayer valley semiconductors, such as tungsten diselenide (WSe$_2$), possess valley pseudospin degrees of freedom that are optically addressable but degenerate in energy. Lifting the energy degeneracy by breaking time-reversal symmetry is vital for valley manipulation. This has been realized by directly applying magnetic fields or via pseudo-magnetic fields generated by intense circularly polarized optical pulses. However, sweeping large magnetic fields is impractical for devices, and the pseudo-magnetic fields are only effective in the presence of ultrafast laser pulses. The recent rise of two-dimensional (2D) magnets unlocks new approaches to control valley physics via van der Waals heterostructure engineering. Here we demonstrate wide continuous tuning of the valley polarization and valley Zeeman splitting with small changes in the laser excitation power in heterostructures formed by monolayer WSe$_2$ and 2D magnetic chromium triiodide (CrI$_3$). The valley manipulation is realized via optical control of the CrI$_3$ magnetization, which tunes the magnetic exchange field over a range of 20 T. Our results reveal a convenient new path towards optical control of valley pseudospins and van der Waals magnetic heterostructures.

The field of valleytronics has flourished with the study of atomically thin transition metal dichalcogenide semiconductors such as $MoS_2$ and $WSe_2$ (ref. 1). Their electrons and holes have inequivalent but degenerate momentum space extrema at the +K and -K conduction and valence band valleys. This valley pseudospin can be initialized in exciton[1–5] populations through circularly polarized optical pumping and manipulated by breaking time-reversal symmetry. For example, the exciton valley splitting, polarization, and coherence are tunable via the Zeeman effect with high magnetic fields[6–13]. It is also possible to coherently control the valleys through the optical Stark effect, which generates large pseudo-magnetic fields that lift the valley degeneracy[14–16] and rotate valley pseudospins[17]. These effective magnetic fields are likely essential for valleytronic devices as well as proposed topological effects unique to monolayer valley semiconductors[18].

The emergence of 2D van der Waals magnetic materials[19–23] expands and enhances the possibilities for valley manipulation via heterostructure engineering[24–28]. In heterostructures of monolayer $WSe_2$ and ultrathin magnetic insulator $CrI_3$, spin-dependent charge hopping across the heterostructure interface leads to excitation-helicity-dependent photoluminescence (PL) intensity[24]. The $CrI_3$ also induces a magnetic exchange field (~13 T) in the $WSe_2$ (Fig. 1a) that generates a sizeable valley Zeeman splitting. These proximity effects enable powerful new means to control monolayer $WSe_2$ valley physics. For instance, valley polarization and valley Zeeman splitting are switchable when the $CrI_3$ magnetization is flipped by an external magnetic field[24]. Here, we demonstrate an optical route to manipulate the interfacial magnetic coupling in $WSe_2/CrI_3$ heterostructures, which enables continuous control of the magnitude and sign of valley polarization and Zeeman splitting.

The heterostructure consists of monolayer $WSe_2$ and ~10 nm $CrI_3$ protectively encapsulated by ~10 to 20 nm hexagonal boron nitride, as shown in the optical microscope image in Fig. 1b. Below its Curie temperature ($T_C$) of 61 K, $CrI_3$ establishes long-range magnetic ordering. The coupling between the out-of-plane $CrI_3$ magnetization and the $WSe_2$ valley pseudospin is measurable from the circularly polarized $WSe_2$ trion PL[24]. All data were taken at 1.6 K under 1.96 eV continuous-wave laser excitation with about a 1 μm excitation spot diameter unless otherwise noted (see Methods).

Figure 1c is a spatial map of the $WSe_2$ PL intensity, which was acquired by rastering the laser over the boxed region in Fig. 1b. By resolving the PL into its circularly polarized components, we can determine the polarization, $\rho = \frac{I_+ - I_-}{I_+ + I_-}$, where $I_\pm$ is the $\sigma^\pm$ PL peak intensity excited by $\sigma^\pm$ polarized laser. Under time-reversal symmetry, the ±K valleys are degenerate and ρ is zero. In proximity to magnetic $CrI_3$, ρ becomes finite and its sign directly correlates with the underlying $CrI_3$ magnetization direction. Spatial maps of ρ thus reveal magnetic domain structure in $CrI_3$. For instance, ρ is spatially uniform and negative (Fig. 1d, right) at 1 T, with representative spectra shown in Fig. 1e. This implies uniform positive magnetization at the interface. However, upon decreasing to 0.7 T (Fig. 1d, left), a large region of the heterostructure flips to positive ρ, signifying a change in the magnetization state of the underlying $CrI_3$. We label this region as domain A, where ρ flips sign three times in a single magnetic field sweep[24], as shown in Fig. 1f. For the remaining top region (which we label as domain B), in contrast, the sign of ρ only flips once (see Supporting Section S1).

Our main finding is that the outer hysteresis loop of domain A near 0.8 T (highlighted in Fig. 1f) strongly depends on the photo-excitation power. In Fig. 2a, ρ is plotted as the magnetic field sweeps from 0.6 to 1.2 T and then back to 0.6 T at selected excitation powers. The hysteresis loop gradually evolves from wide and square at 1 μW to narrow and sloped at 100 μW. This photoinduced change in the coercivity has dramatic consequences for the optical control of valley properties at fixed magnetic fields near the hysteresis loop. We performed power-dependent measurements of ρ at fixed magnetic fields from 0.7 to 1 T, as shown in the 2D plot of ρ in Fig. 2b. The magnetic field was first initialized by sweeping up from 0.6 T. Below ~0.75 T and above ~0.92 T, ρ decreases in magnitude slightly with increasing power, but its sign remains the same. In striking contrast, at intermediate fields, the sign of ρ flips at high powers. The curved white area shows that the critical power decreases at higher magnetic fields, consistent with the power dependence of the hysteresis loop coercivity in Fig. 2a. The vertical line in Fig. 2a shows how the power-dependent coercivity can cause the sign flip in ρ at a fixed magnetic field. We also find similar power dependence and switching behavior for the valley Zeeman splitting (Δ), which was extracted by the energy difference between $\sigma^{\pm}$ PL peaks (Section S2).

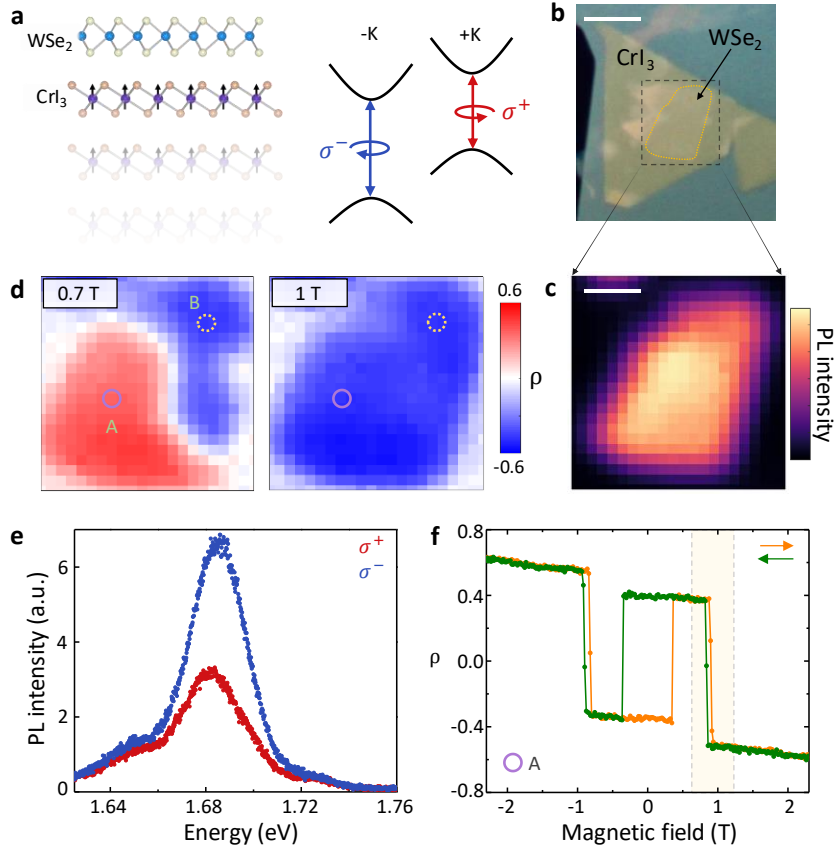

**Figure 1. Basic characterization and domains of WSe$_2$/CrI$_3$ heterostructure.** (**a**) Schematic of WSe$_2$/CrI$_3$ heterostructure (left). Valley energy level diagram and optical selection rules of monolayer WSe$_2$ with magnetic exchange field coupling (right). h-BN encapsulation layers are not shown. (**b**) Optical microscope image of heterostructure. Dashed box region shows the laser scanning area and the dotted yellow curve outlines the WSe$_2$ monolayer region. Scale bar, 5 μm. (**c**) Spatial map of total photoluminescence (PL) intensity within the boxed region of Fig. 1b. Scale bar, 2 μm. (**d**) Spatial maps of the polarization parameter ρ (see text for definition) at 1 T (right) and 0.7 T (left) applied magnetic field. Same spatial scale as Fig. 1c. (**e**) Spectra of $\sigma^+$ ($\sigma^-$) PL under $\sigma^+$ ($\sigma^-$) laser excitation taken at 1 T applied magnetic field, shown in red (blue). (**f**) Magnetic field dependence of ρ for up (orange) and down (green) field sweep directions. The data was taken on domain A at the location marked by the solid purple circle in Fig. 1d. The corresponding data for domain B (marked by dashed yellow circle) is in Section S1.

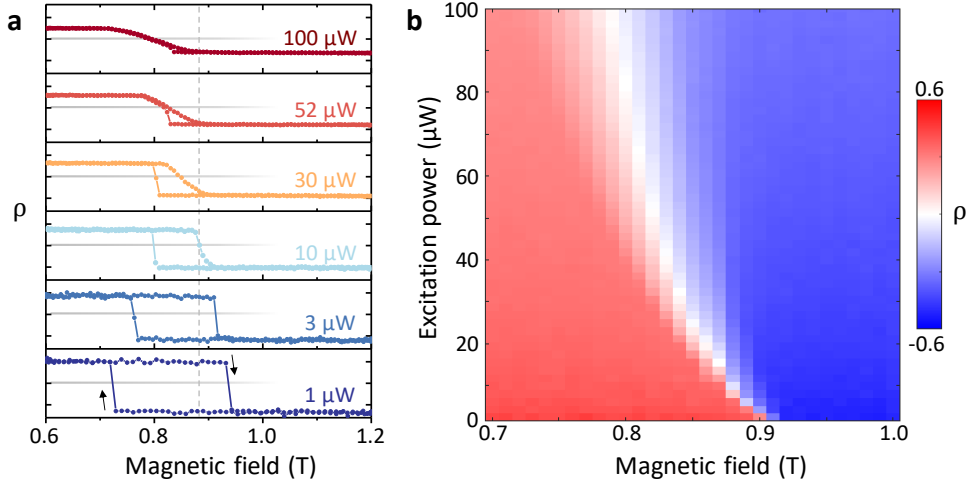

**Figure 2**. **Power-dependent hysteresis and valley switching.** (**a**) Magnetic field sweeps of ρ from 0.6 to 1.2 to 0.6 T (highlighted region in Fig. 1f) at different excitation powers. Gray horizontal lines indicate ρ = 0 and the neighboring tick marks on the y-axis are ±0.5. Sweep directions are shown by the black arrows. (**b**) Map of the power dependence of ρ taken at different applied magnetic fields. The magnetic field was first initialized by sweeping up to 0.7 T. A power dependence was then performed from 1 to 100 μW at the different fixed magnetic fields from 0.7 to 1 T.

The power-switchable valley properties are clearly illustrated in the PL spectra in Fig. 3a taken at 0.88 T. At low excitation powers, $I_+$ (red curve) is more intense and has higher energy than $I_-$ (blue curve). With increasing power, however, $I_+$ and $I_-$ become degenerate and eventually they switch in their relative intensity and energy. From 4 to 40 μW, ρ continuously changes from 0.41 to -0.37 (Fig. 3b) and Δ from 3.7 to -1.3 meV (Fig. 3c, also Section S3). To produce comparable switching of the valley splitting with a bare WSe$_2$ monolayer would require sweeping an external magnetic field between -15 and 5 T. These results are consistent with a second heterostructure sample (Section S4). We provide additional power dependences for ρ and Δ at different magnetic field values in Section S5. The valley switching effects were observed for both circular and linear excitation polarization, and therefore only the total optical excitation power matters (Section S6). We also find that the power dependence is identical with both increasing and decreasing power, which indicates the effects are fully reversible with no hysteresis (Fig. 3b, Section S7). These results demonstrate reversible optical control of the valley polarization (between about ±40%) and magnetic exchange field (Fig. 3c), by varying the excitation power within an order of magnitude.

As revealed from the power-dependent spatial maps of ρ in Fig. 3c, the reversible valley switching occurs on all areas of domain A. The arrows denote the acquisition order of the maps, which were all taken at 0.84 T. At low laser excitation power, there are two areas of opposite polarization (see 10 μW plot in Fig. 3c), which correspond to the two magnetic domains as discussed. When the excitation power increases to 100 μW, the domain of positive polarization (A) completely reverses, which implies that optical excitation can flip all areas of domain A. After the excitation power is lowered back to 10 μW, the original domain pattern recovers.

The valley switching effect arises from optical control of the $CrI_3$ magnetization and hence the resulting magnetic proximity effects. To further unravel the connection, we directly probed the $CrI_3$ magnetization via reflection magnetic circular dichroism (RMCD) on domain A. RMCD measures the difference in reflection between right and left circularly polarized light and is proportional to the total out-of-plane magnetization in the $CrI_3$. In Fig. 4a, we show the magnetic field dependence of the RMCD signal at different excitation powers, which should be compared to the study for ρ from Fig. 2a. The RMCD exhibits a very similar power-dependent hysteresis loop behavior to ρ; the full-width of the RMCD loop decreases and the loop slants with increasing power. However, unlike with ρ, the RMCD does not switch signs. Further differences are revealed in Fig. 4b, where the RMCD is tracked over a wider magnetic field range. In contrast to ρ, which flips three times in a magnetic field sweep (Fig. 1f), the RMCD signal contains many step-like jumps and is monotonic with the magnetic field (Fig. 4b). While there are changes in the RMCD signal where ρ and Δ flip sign (near ±0.8 T), the additional jumps near 0 and ±1.9 T do not appear in the behavior of ρ. The RMCD data thus reveal that the $CrI_3$ magnetization changes discretely in several steps, unlike a typical ferromagnet.

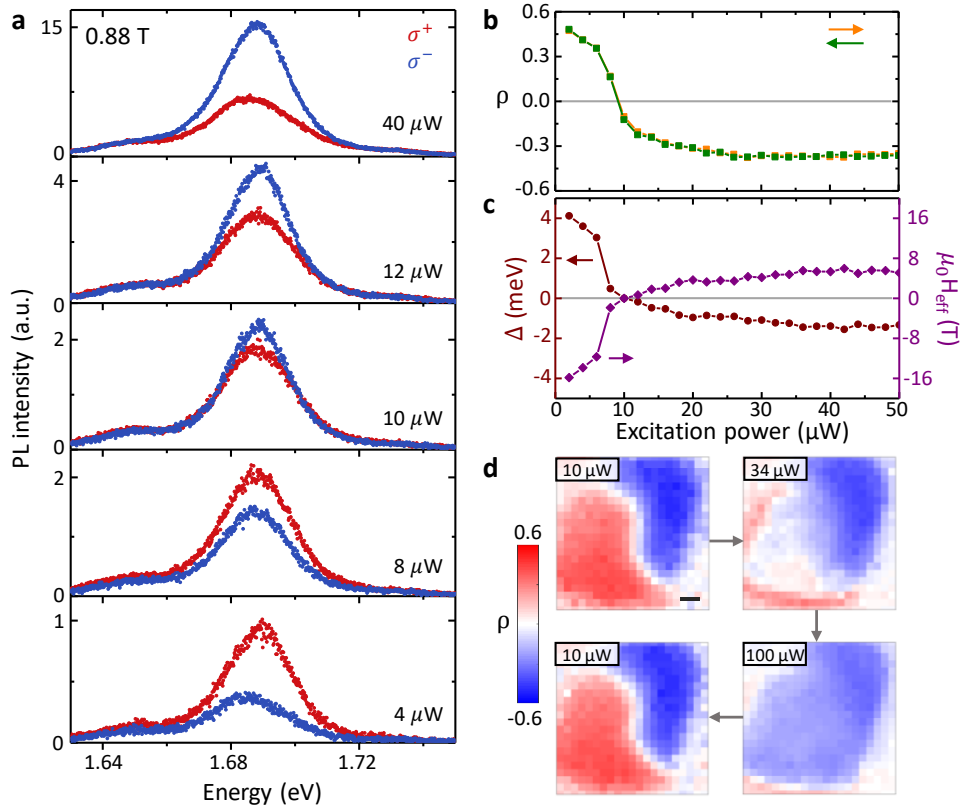

**Figure 3. Manipulation of valley polarization and splitting via optical excitation power.** (**a**) Circularly polarized PL spectra at selected excitation powers. The applied magnetic field was initialized to 0.88 T from 0.6 T. (**b**) Power dependence of ρ at 0.88 T with increasing (orange) and decreasing (green) power. (**c**) Power dependence at 0.88 T of the valley splitting (Δ, left) and the corresponding effective magnetic field ($\mu_0 H_{eff}$, right). The full data for Fig. 3b and c are in Section S3. (**d**) Spatial maps of ρ at 10, 34, 100, and 10 μW, in that order. The black scale bar represents the laser beam diameter (1 μm), which is much smaller than the domain size.

We can begin to understand this behavior if we assume that the $WSe_2$ monolayer is primarily influenced by the top $CrI_3$ layer. This assumption is reasonable since the exchange coupling is short-ranged, so the magnetic interactions between $WSe_2$ and deeper $CrI_3$ layers are suppressed. In fact, the data imply that the changes in ρ at the 0.8 T hysteresis loop are caused by magnetization reversal within a single layer of $CrI_3$. From atomic force microscopy of our $CrI_3$ flake, we know its thickness is about 10 nm, or 15 to 16 layers thick (~0.66 nm per layer)[23]. As shown in Fig. 4b, near 0.8 T, the RMCD signal jumps by ~0.013, or about 1/8th of the saturation value at 2.3 T (RMCD ~ 0.103). Therefore, the $CrI_3$ magnetization also changes by ~1/8, implying that 1/16th of the $CrI_3$ reverses its magnetization due to its layered antiferromagnetic nature[19]. Given the $CrI_3$ thickness (~16 layers), it strongly suggests that the magnetization step at 0.8 T occurs due to the flipping of a single $CrI_3$ layer. We thus establish that the valley switching behavior near 0.8 T originates from the photoinduced flipping of a single interfacial layer of $CrI_3$ magnetization. In addition, the other step-like jumps in the RMCD sweeps (Fig. 4b) must arise from magnetization changes within deeper layers and do not impact the $WSe_2$ to the same extent. We can also now refine our understanding of the differences between domain A and B. In a magnetic field sweep, the topmost $CrI_3$ layer flips three times in domain A, but only once in domain B (Section S1). Near 0.8 T in domain B, only deeper layers away from the $WSe_2$ interface are susceptible to flipping, which explains the negligible power-dependent PL polarization.

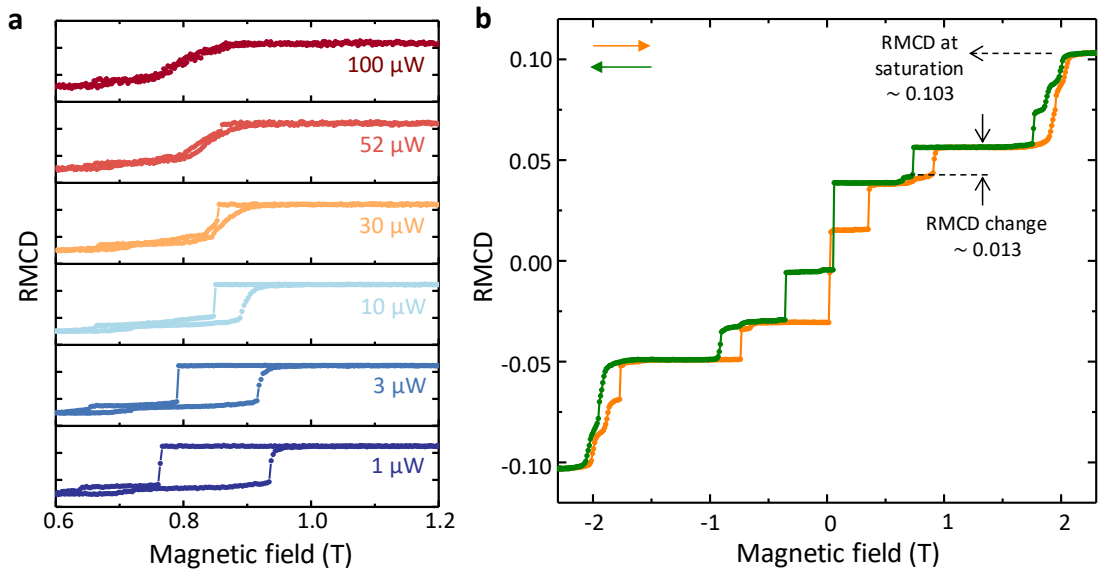

**Figure 4. $CrI_3$ magnetization and temperature dependence of hysteresis loop.** (**a**) Magnetic field dependence of the RMCD from 0.6 to 1.2 to 0.6 T at selected excitation powers. The RMCD range in each box is from 0.03 to 0.06 and the tick marks are separated by 0.01. (**b**) Magnetic field dependence of RMCD sweeping up (orange curve) and down (green curve) on domain A at 10 μW optical excitation power.

Since the top CrI$_3$ layer predominantly impacts the WSe$_2$, one might expect that the magnetic exchange field should be equal and opposite after flipping the top CrI$_3$ layer. However, from Fig. 3c, the magnitude of the effective exchange field at high powers (~5 T) is less than the magnitude at the lowest powers (~15 T). The origin of this discrepancy is currently unclear. One possibility is due to CrI$_3$ demagnetization at high powers, since small magnetization changes in the topmost CrI$_3$ layer can have an outsized impact on the WSe$_2$ layer. Another possibility is that the photoexcited excitons and free carriers at higher powers influence the valley depolarization in WSe$_2$. Future studies should consider pump-probe measurements, with pump light below the WSe$_2$ optical gap to control CrI$_3$ and weak above-gap excitation to measure the WSe$_2$ PL. Such experiments will provide insight into the excitation energy dependence of the opto-magnetic effects and help to disentangle the WSe$_2$ and CrI$_3$ power dependences.

Another essential topic for further study is the underlying mechanism for the photoinduced changes in the CrI$_3$ layer magnetization. While optically controlled magnetism is a vast field of research, the optical control of magnetic coercivity using relatively low-power continuous-wave excitation has only been observed in a few magnetic semiconductor systems, such as (Ga,Mn)As (ref. 29), Ni/GaAs (ref. 30), and (In,Mn)As/GaSb (ref. 31). In these samples, photoexcited carriers reduce the coercivity by enhancing the carrier-mediated exchange interactions, which lower the domain wall energies. For CrI$_3$, our preliminary measurements show that width of the 0.8 T hysteresis loop is very sensitive to temperature, which suggests that laser heating of the lattice may be an important factor (Section S8). We also observe that outermost RMCD hysteresis loops at ±1.9 T are sensitive to power (Section S9). Therefore, excitation power also impacts deeper CrI$_3$ layers, and the photo-induced layer flipping effect can likely occur in CrI$_3$ in the absence of WSe$_2$. We emphasize that further theoretical studies and detailed experiments, such as gate dependence, CrI$_3$ thickness dependence, and time-resolved studies, will be required to fully understand the opto-magnetic effects in CrI$_3$, which are beyond the scope this work.

In conclusion, we have demonstrated a new route to manipulate the WSe$_2$ valley pseudospins by optical control of the magnetic proximity effect. Using small changes in laser power, we can reversibly flip the top CrI$_3$ layer magnetization, which tunes the magnetic exchange field over a range of 20 T and thus controls the valley polarization and Zeeman splitting without changing the external magnetic field. These observations are uniquely enabled by the ability to fabricate a high-quality van der Waals heterostructure between a 2D magnetic insulator with large domain size and a non-magnetic monolayer valley semiconductor. A clean heterostructure interface, together with optically sensitive magnetic properties, is challenging to realize using conventional magnetic insulators. The optically tunable magnetic exchange field demonstrated here should be generalizable to a wide variety of CrI$_3$-based van der Waals heterostructures, which may be a powerful tool in the study of physical phenomena requiring time-reversal symmetry breaking.

## Methods

### Device fabrication

Bulk crystals of WSe$_2$, CrI$_3$, and h-BN were first exfoliated onto 90 nm SiO$_2$ on Si in a glovebox with O$_2$ and H$_2$O levels below 0.5 ppm. After finding the CrI$_3$ sample, we used a polycarbonate-based transfer technique to assemble the heterostructure (h-BN/CrI$_3$/monolayer WSe$_2$/h-BN) in the glovebox[24]. The chloroform rinse was performed in ambient environment for 2 minutes. The h-BN thickness was ~ 10 to 20 nm and the CrI$_3$ thickness was ~10 nm.

### Optical measurements

The samples were measured in reflection geometry using a dry cryostat equipped with a 9 T superconducting magnet in Faraday configuration. Continuous-wave excitation from a power-stabilized laser at 1.96 eV was focused to ~1 μm$^2$ with an aspheric lens, and the collected PL was detected with a spectrometer and Si charge-coupled device. The excitation polarization and power, as well as the detection polarization, were controlled by liquid-crystal variable waveplates and linear polarizers. For the excitation power sweeps, an integration time of 1 second was used for each polarization, and subsequent power data points were acquired within 2 seconds of each other. For the magnetic field sweeps, the sweep rate was ~7 mT/s and the PL integration time was 2 seconds for the 1 μW curves and 1 second for all other curves. Spatially resolved measurements were performed by scanning the sample position using a piezoelectric stage. The PL peak intensity and energy were extracted from a double pseudo-Voigt function fit to the trion and defect peaks[24]. The effective magnetic field was calculated from the valley splitting, assuming a $g$-factor of -4.5 (ref. 24). RMCD was measured using the same laser with a photoelastic modulator as detailed in ref. 32.

**Acknowledgments:** This work is mainly supported by the Department of Energy, Basic Energy Sciences, Materials Sciences and Engineering Division (DE- SC0018171). Device fabrication is partially supported by University of Washington Innovation Award. WY is supported by the Croucher Foundation (Croucher Innovation Award) and the HKU ORA. Work at ORNL (MAM) was supported by the US Department of Energy, Office of Science, Basic Energy Sciences, Materials Sciences and Engineering Division. KW and TT acknowledge support from the Elemental Strategy Initiative conducted by the MEXT, Japan and a Grant-in-Aid for Scientific Research on Innovative Areas "Science of Atomic Layers" from JSPS. DX and KF acknowledge the support a Cottrell Scholar Award. XX acknowledges the support from the State of Washington funded Clean Energy Institute and from the Boeing Distinguished Professorship in Physics.

**Author Contributions:** XX and KLS conceived the experiment. DZ and BH fabricated the samples. KLS acquired the experimental data, assisted by DZ, XL, and NPW, supervised by XX and KCF. KLS analyzed the data. MAM synthesized and characterized the bulk CrI$_3$ crystal. TT and KW synthesized and characterized the bulk boron nitride crystal. DX and WY provided theoretical support. KLS and XX wrote the manuscript with input from all authors. All authors discussed the results.

**Competing Financial Interests:** The authors declare no competing financial interests.

**Supporting Information:** The following supporting information is available. Domain B field dependence; full valley Zeeman splitting data; power-dependent PL parameters; data from additional sample; additional power dependence at different fields; linear versus circular excitation; influence of magnetic field initialization direction; temperature dependence; wide field dependence at different powers.

# Supporting Information for

# Valley Manipulation by Optically Tuning the Magnetic Proximity Effect in WSe$_2$/CrI$_3$ Heterostructures

Kyle L. Seyler[1], Ding Zhong[1], Bevin Huang[1], Xiayu Linpeng[1], Nathan P. Wilson[1], Takashi Taniguchi[2], Kenji Watanabe[2], Wang Yao[3], Di Xiao[4], Michael A. McGuire[5], Kai-Mei C. Fu[1,6], Xiaodong Xu[1,7*]

[1]Department of Physics, University of Washington, Seattle, Washington 98195, USA
[2]National Institute for Materials Science, 1-1 Namiki, Tsukuba 305-0044, Japan
[3]Department of Physics and Center of Theoretical and Computational Physics, University of Hong Kong, Hong Kong, China
[4]Department of Physics, Carnegie Mellon University, Pittsburg, Pennsylvania 15213, USA
[5]Materials Science and Technology Division, Oak Ridge National Laboratory, Oak Ridge, Tennessee, 37831, USA
[6]Department of Electrical Engineering, University of Washington, Seattle, Washington 98195, USA
[7]Department of Materials Science and Engineering, University of Washington, Seattle, Washington 98195, USA

*Corresponding author. Email: xuxd@uw.edu. Phone: (206) 543-8444

**Contents:**

S1. Magnetic field dependence of domain B

S2. Power-dependent hysteresis and switching of the valley Zeeman splitting

S3. Raw power-dependent PL parameters at 0.88 T

S4. Power-dependent hysteresis and valley switching on a second sample

S5. Power dependence of ρ and Δ at selected magnetic fields

S6. Comparison between linear and circular excitation

S7. Magnetic field initialization direction and partial hysteresis

S8. Temperature dependence

S9. Wide magnetic field sweeps at different powers

## S1. Magnetic field dependence of domain B

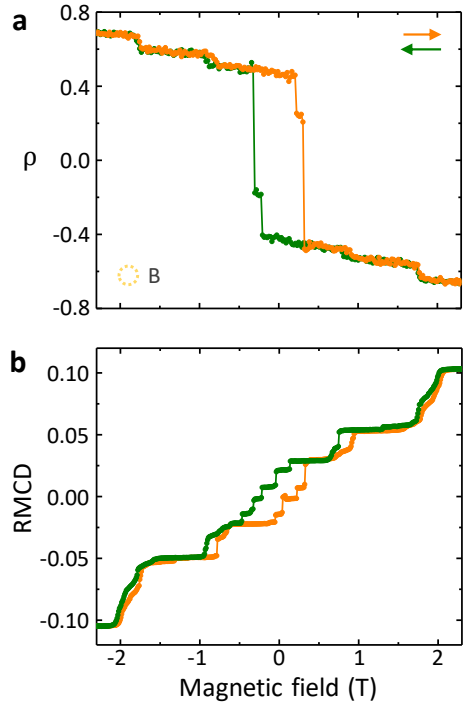

**Figure S1** | Magnetic field dependence of ρ (**a**) and RMCD (**b**) sweeping up (orange curves) and down (green curves) on domain B. The data are taken at the location indicated by the dashed yellow circle in Fig. 1d.

In Fig. S1, ρ (Fig. S1a) and RMCD (Fig. S1b) are plotted as a function of the applied magnetic field on domain B. The sign of ρ flips only once in each sweep direction, whereas in domain A, ρ flips sign three times. The magnetic field dependence of the RMCD is monotonic with many small step-like jumps. These jumps are a sign of magnetization changes within individual layers of the ~10 nm $CrI_3$. Since the $WSe_2$ valley physics is primarily affected by the topmost $CrI_3$ layer (see further discussion in the main text), we infer that in domain B, the top layer of $CrI_3$ only flips magnetization once, which leads to single sign flip of ρ in Fig. S1a. The spin-orientation-dependent charge transfer from $WSe_2$ to the $CrI_3$ leads to an opposite sign for the RMCD signal compared to ρ at high magnetic fields. For example, at high positive magnetic field, the magnetization (and RMCD signal) is positive, which opens access to charge transfer from $|K,\uparrow\rangle$ electrons from $WSe_2$, quenches the $\sigma^+$ PL, and thus generates negative ρ.

## S2. Power-dependent hysteresis and switching of the valley Zeeman splitting

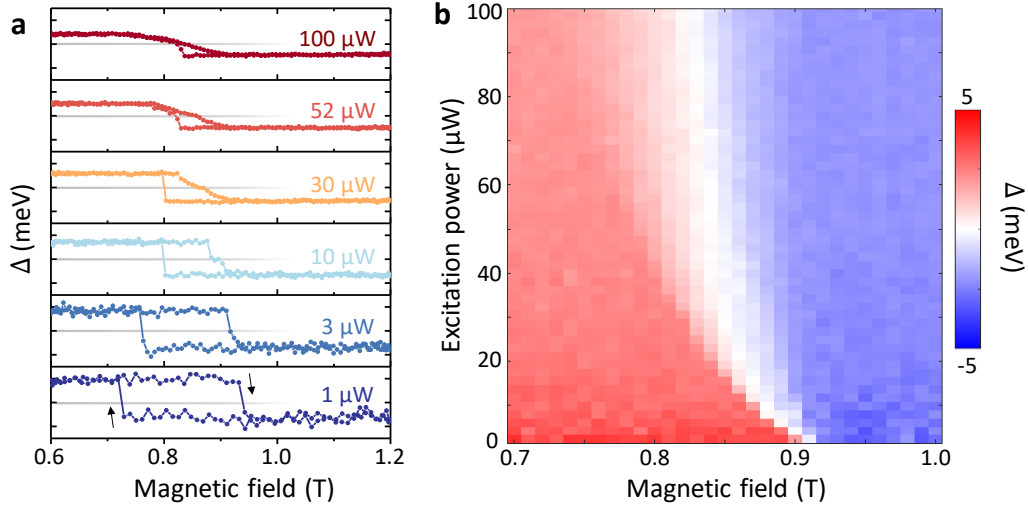

**Figure S2 |** (**a**) Magnetic field dependence of valley Zeeman splitting (Δ) from 0.6 to 1.2 to 0.6 T. The gray horizontal lines indicate Δ = 0 and the neighboring y-axis tick marks denote Δ = ±4 meV. Black arrows indicate the sweep directions. These data are extracted from the same spectra used in Fig. 2a of the main text. (**b**) Map of the power dependence at different applied magnetic fields for Δ. The magnetic field was first initialized by sweeping up to 0.7 T. A power dependence was then performed from 1 to 100 μW at the different fixed magnetic fields from 0.7 to 1 T. These are extracted from the same dataset used in Fig. 2b from the main text.

Figure S2 provides similar data to Fig. 2 in the main text for the valley Zeeman splitting (Δ). It is clear that Δ exhibits the same power-dependent hysteresis loop as ρ (Fig. S2a). We are thus able to control Δ with the optical excitation power at fixed magnetic fields near the hysteresis loop, as shown in the 2D plot of Δ in Fig. S2b. The sign of the valley Zeeman splitting is tunable from positive to negative (and the reverse) when the external magnetic field is between ~0.82 T and 0.9 T. The curved white region indicates the critical excitation power that is required to switch the sign of Δ at the different fixed magnetic fields.

## S3. Raw power-dependent PL parameters at 0.88 T

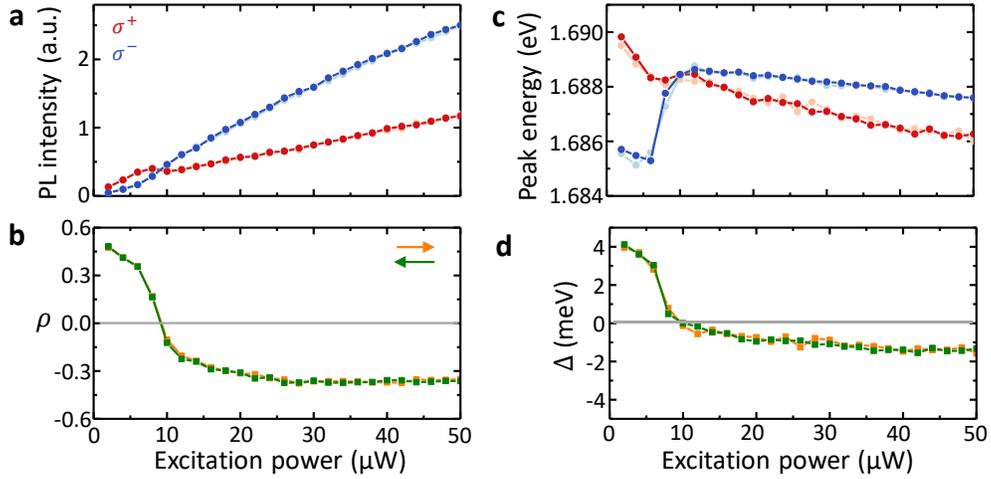

**Figure S3** | Power dependence of (**a**) the PL intensity and (**c**) peak energy for $\sigma^+$ (red) and $\sigma^-$ (blue) excitation and detection polarization. Light (dark) red and blue show the values for increasing (decreasing) power. The corresponding valley parameters ρ and Δ are shown in (**b**) and (**d**) respectively for increasing (orange) and decreasing (green) power. The power was swept as high as 100 μW, but we display it up to 50 μW for easier viewing of the low-power nonlinearity. Before the power dependence, the external magnetic field was initialized to 0.88 T by sweeping up from 0.6 T. See Fig. S5 for the full power dependence of ρ and Δ.

Figures S3 demonstrates the strong nonlinearity of the PL peak parameters that produces the valley switching effects. For example, in Fig. S3a, the $\sigma^+$ PL intensity power dependence exhibits a kink and its slope decreases above 8 μW. This causes the $\sigma^+$ and $\sigma^-$ curves to intersect at ~10 μW, which leads to a reversal in the sign of ρ. There is good overlap of the data when the power dependence is performed with increasing and then decreasing power, which shows the reversibility of the phenomenon.

## S4. Power-dependent hysteresis and valley switching on a second sample

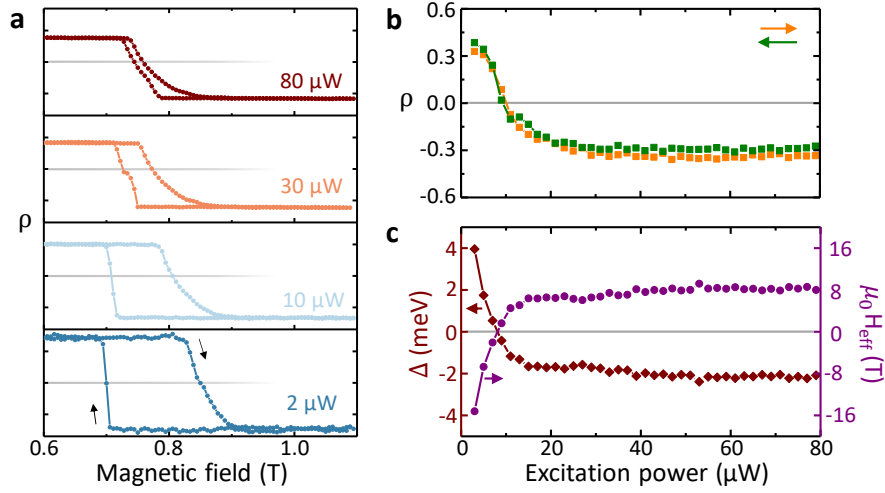

**Figure S4 |** (**a**) Magnetic field sweeps of ρ from 0.6 to 1.1 to 0.6 T at selected excitation powers on a second WSe$_2$/CrI$_3$ sample. Gray horizontal lines indicate the ρ = 0 line and the neighboring tick marks on the y-axis are ±0.5. Sweep directions are shown by the black arrows. (**b**) Power dependence of ρ at 0.78 T from 3 to 80 µW (orange curve) and subsequently from 80 to 3 µW (green curve). (**c**) Power dependence of Δ (left) at 0.78 T and the corresponding effective magnetic field ($\mu_0 H_{eff}$, right). All data from the second device were acquired at 8 K.

In Fig. S4, we provide additional data from a second heterostructure sample of WSe$_2$/CrI$_3$. The measurements were acquired at 8 K on a domain where ρ flips three times in a single magnetic field sweep, like domain A in the sample from the main text. The hysteresis loop at 0.8 T exhibits very strong power-dependent coercivity and loop shape (Fig. S4a), just as with the other sample. We replicate the valley switching effects by fixing the magnetic field at 0.78 T and varying the excitation power. We achieve reversible switching between ±30% polarization (Fig. S4b) and around 4 to -2 meV valley Zeeman splitting (Fig. S4c). This change in valley Zeeman splitting corresponds to varying the effective magnetic exchange field from about -15 T to 7 T, via a modest increase in the laser excitation power.

## S5. Power dependence of ρ and Δ at selected magnetic fields

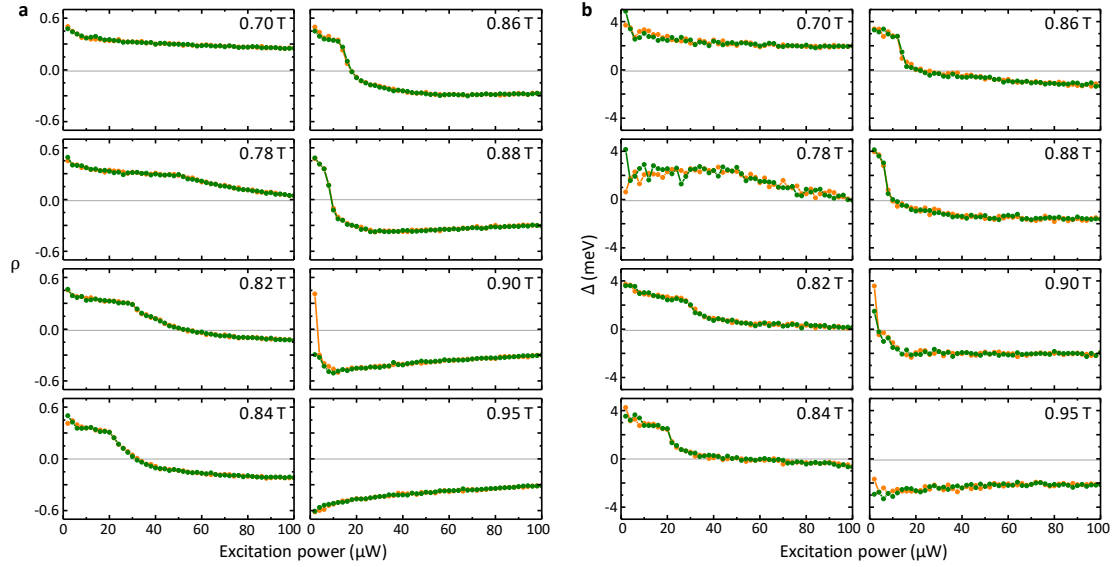

**Figure S5.** Power dependence of ρ (**a**) and Δ (**b**) at selected applied magnetic fields with increasing (orange) and decreasing (green) power.

In Fig. S5, we display several power dependences for ρ (Fig. S5a) and Δ (Fig. S5b) at different (fixed) applied magnetic fields. They correspond to vertical line cuts in the 2D maps Fig. 2b and S2b. The indicated magnetic field is first initialized by sweeping up from 0.6 T. Below 0.80 T and above 0.90 T, the signs of ρ and Δ do not change with power. At intermediate magnetic fields, both ρ and Δ flip when the excitation laser reaches a critical power that is determined by the magnetic field. Higher magnetic fields decrease the critical power required for valley switching, but around 0.90 T and above, the system tends towards the high magnetic field state and thus the reversible power dependence is less robust.

## S6. Comparison between linear and circular excitation

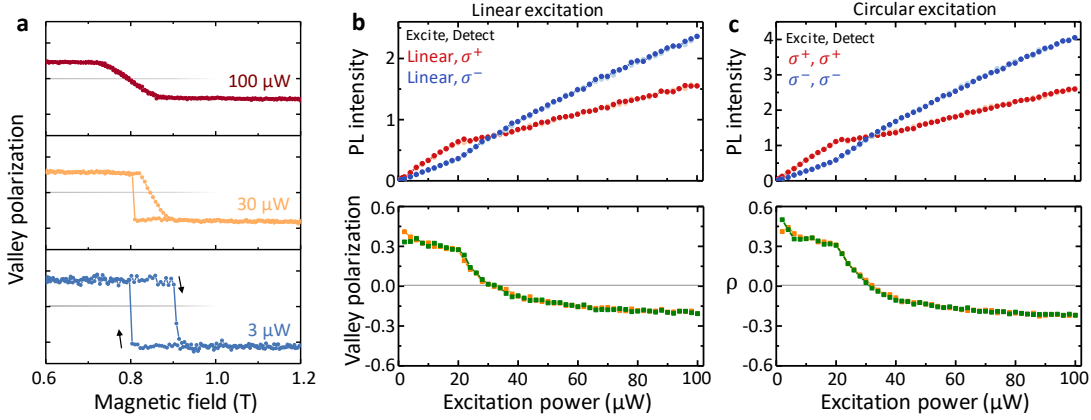

**Figure S6** | (**a**) Magnetic field dependence of the valley polarization (circular polarization under linear excitation) at selected excitation powers from 0.6 to 1.2 to 0.6 T. Gray horizontal lines indicate the $\rho = 0$ line and the neighboring tick marks on the y-axis are ±0.5. Sweep directions are shown by the black arrows. Power dependence of PL intensity and $\rho$ for linear (**b**) and circular excitation (**c**) at 0.84 T.

Figure S6 shows that the valley switching effect occurs under both linear and circular excitation. The linear excitation, circular detection condition allows us to detect the valley polarization that is induced by the magnetic proximity effect. As seen in Fig. S6a, the valley polarization exhibits very similar power-dependent hysteresis behavior as $\rho$ in Fig. 2a, including decreased full-width and increased transition width (slanted region where $\rho$ reverses) at high powers. We also performed a power dependence under linear (Fig. S6b) and circular (Fig. S6c) excitation at 0.84 T. Aside from a ~5% enhancement, $\rho$ behaves the same as valley polarization. From $\rho$, we can thus infer the true valley polarization, as reported in the main text. We note that measuring $\rho$ has the benefit of higher PL intensities, which decreases the data integration time required for high-quality spectra. Also, by using circular instead of linear excitation, we avoid the complication of a rotating excitation linear polarization angle with the magnetic field (due to Faraday rotation). Thus, the magnetic field dependence of $\rho$ is robust against possible sample anisotropies. A more detailed exploration of linear polarization (excitation and detection) should be performed in the future. Nevertheless, the power-dependent effects we observe here are independent of the excitation polarization and rely only on the total power.

## S7. Magnetic field initialization direction and partial hysteresis

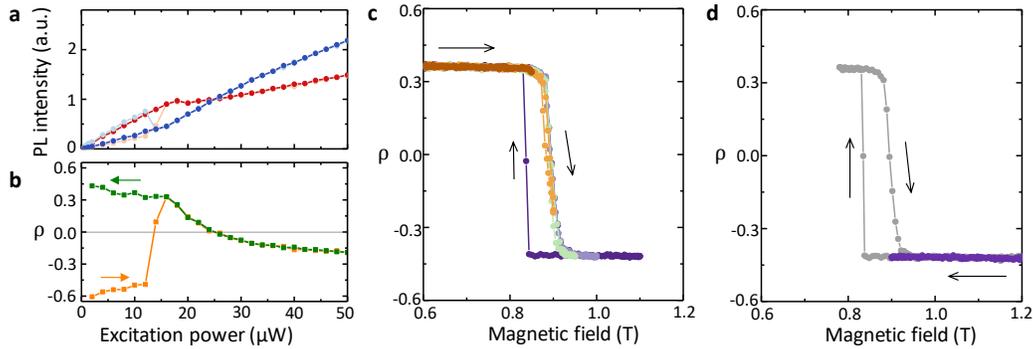

**Figure S7** | Power dependence of the PL intensity (**a**) and ρ (**b**) after initializing the applied magnetic field to 0.85 T by sweeping *down* from 1.2 T. Light (dark) red and blue curves in (a) show the PL intensity for increasing (decreasing) power. (**c**) Partial magnetic field sweeps from 0.6 T to different maximum fields and back to 0.6 T. (**d**) Partial magnetic field sweep from 1.2 T to 0.9 T to 1.2 T (purple) and 1.2 T to 0.78 T to 1.2 T (gray).

The magnetic field initialization direction affects the power dependence and hysteresis. Figures S7a and b are power dependences of circularly polarized PL intensities and ρ taken at 0.85 T. Unlike the other power dependences that we show, however, here we initialized the magnetic field by sweeping *down* from 1.2 T. We thus begin our power dependence in the high-magnetic-field state (which has negative ρ). We first explored the effect of increasing power (orange curve in Fig. S7b). At ~14 µW, ρ jumps sharply from negative to positive, and then slowly decreases in magnitude at higher power, eventually flipping back to negative above 25 µW. Upon performing the same power dependence in the reverse direction (decreasing power, green curve), we find a curve very similar to those in Fig. S5, and there is a mismatch with the orange curve at 14 µW and below. The jump we observe when increasing from 12 to 16 µW reflects a sudden transition from the high-magnetic-field state to the low-magnetic-field state and indicates that the high-field state is unstable near the hysteresis loop.

Further evidence of the high-field and low-field state stabilities are observed in the partial magnetic field sweeps of ρ. Figure S7c shows several partial field sweeps where the magnetic field starts at 0.6 T, increases to different maxima, and then decreases back to 0.6 T. Even after ρ has flipped from positive to negative (light purple, green, and orange curves), the sweep down curves closely retrace the sweep up curves with low hysteresis, and the system remains in the low-field state. To observe the full hysteresis loop, the magnetic field must sweep up to over 1.1 T, as shown in the dark purple curve. In contrast, if the partial field sweeps begin in the high-field state at 1.2 T (Fig. S7d), the system exhibits the full hysteresis as soon as ρ flips sign (gray curve). These data demonstrate the strong stability of the low-field state compared to the high-field state and reveal why the power-dependent valley switching is fully reversible when the system is initialized from low magnetic fields.

## S8. Temperature dependence

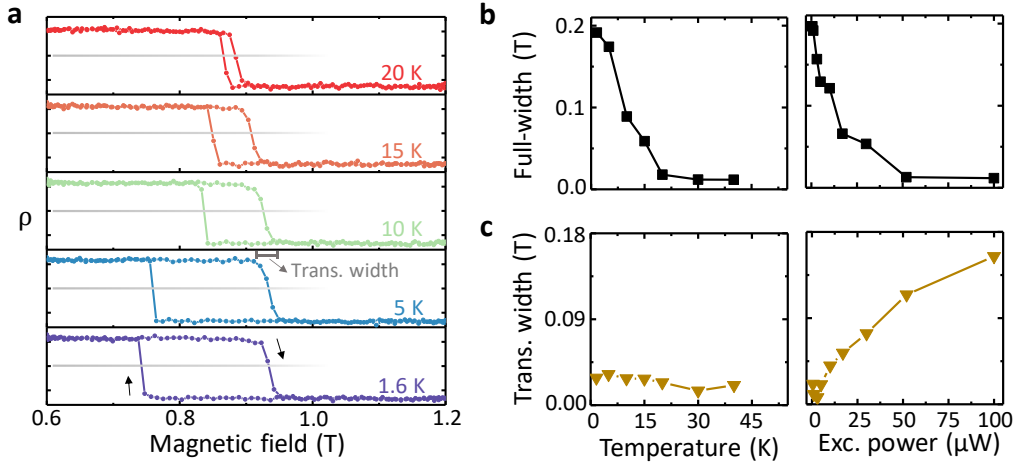

**Figure S8.** (**a**) Magnetic field sweeps of ρ as a function of temperature. The excitation power is 1 μW. Horizontal gray lines indicate ρ = 0 and the neighboring tick marks on the y-axis are ±0.5. Black arrows indicate the field sweep directions. (**b**) Full-width of the hysteresis loop (at ρ = 0) versus temperature at 1 μW excitation power (left) and versus excitation power at 1.6 K (right). (**c**) Transition width versus temperature for 1 μW power (left) and versus excitation power at 1.6 K. See (a) for definition of the transition width.

As discussed in the main text, the origin of the optical control of the CrI$_3$ layer magnetization is an interesting issue for further study. A possible contribution is laser heating of the lattice. As a preliminary test for heating effects, we measured the hysteresis loop of ρ under low power excitation (1 μW) as a function of temperature, as shown in Fig. S8a. The full-width of the loop decreases with temperature, similar to what occurs at high powers (Fig. S8b). On the other hand, there is a clear difference in the shape of hysteresis loop and the magnetic field range of the transition where ρ reverses (i.e., transition width). From 1.6 to 40 K, the transition width remains small (Fig. S8c), as is clear from the box-like loop shapes in Fig. S8a. In contrast, higher power forces a slant in the loop shape (Fig. 2a), increasing the transition width. The magnetization thus appears to gradually rotate with the magnetic field at high powers, unlike the rapid spin-flip transition that occurs at lower powers. However, an important point to emphasize is that the focused laser spot (~1 μm$^2$) is significantly smaller than the sample. Therefore, in comparing the temperature and power dependences, we must consider the difference between local inhomogeneous laser-induced heating and the global homogeneous effects of higher bath temperature. In addition, the inhomogeneous distribution of photoexcited carriers from the laser spatial profile can be relevant, as it may provide a spatially dependent magnetic anisotropy across the laser spot that affects the hysteresis behavior. Overall, the evidence suggests that laser heating is an important effect to consider in future analyses. Systematic exploration of these issues as well as other potential opto-magnetic effects should be undertaken in future experiments and theoretical studies.

## S9. Wide magnetic field sweeps at different powers

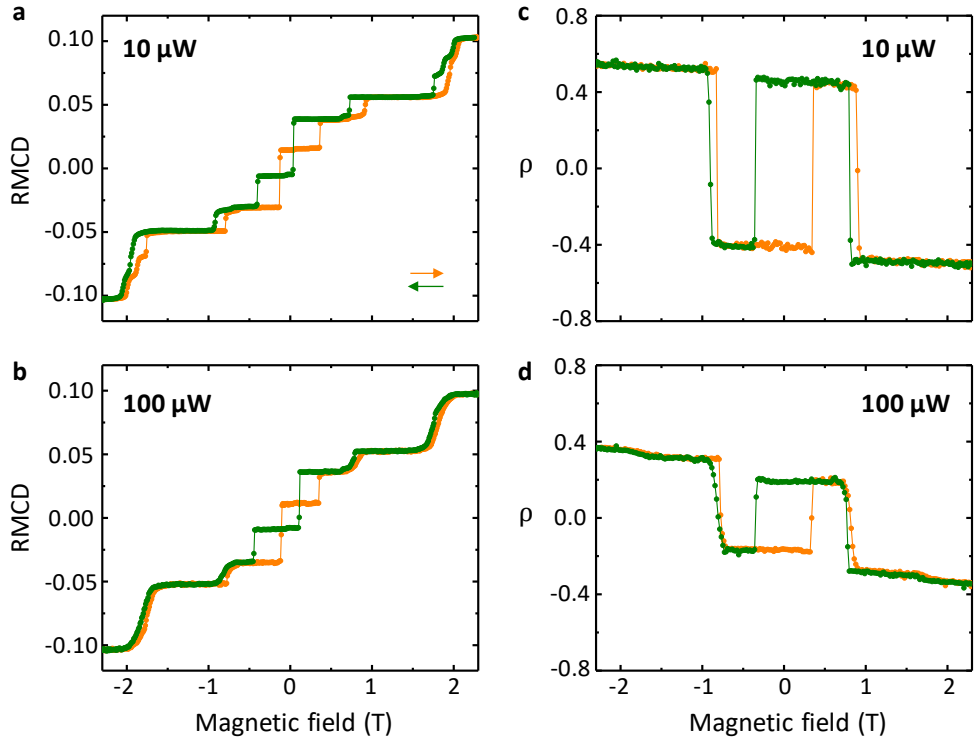

**Figure S9.** Magnetic field sweeps of RMCD and ρ at different excitation powers. RMCD at 10 μW (**a**) and 100 μW (**b**). ρ at 10 μW (**c**) and 100 μW (**d**). In the RMCD sweeps, the hysteresis loops at ±0.8 T and ±1.9 T shrink at high power, while the other loops around zero field are largely unchanged.